# Design and Implementation of Flutter based Multi-platform Docker Controller App

Adarsh Saxena
*Department of Electronics and Communication*
*University of Allahabad*
Prayagraj, India
adarshsaxena358@gmail.com

Sudhakar Singh
*Department of Electronics and Communication*
*University of Allahabad*
Prayagraj, India
sudhakar@allduniv.ac.in

Shiv Prakash
*Department of Electronics and Communication*
*University of Allahabad*
Prayagraj, India
shivprakash@allduniv.ac.in

Nand Lal Yadav
*Department of Electronics and Communication*
*University of Allahabad*
Prayagraj, India
nly.jct@gmail.com

Tiansheng Yang
*Faculty of Business and Creative Industries*
*University of South Wales*
Pontypridd, United Kingdom
tiansheng.yang1@southwales.ac.uk

Rajkumar Singh Rathore
*Cardiff School of Technologies*
*Cardiff Metropolitan University*
Cardiff, United Kingdom
rsrathore@cardiffmet.ac.uk

Shreya Singh
*Department of Computer Science & Engineering*
*Institute of Engineering and Technology,*
*Dr. A.P.J. Abdul Kalam Technical University*
Lucknow, India
shreyasingh20001126@gmail.com

*Abstract*—This paper focuses on developing a Flutter application for controlling Docker resources remotely. The application provides a user-friendly interface for executing various Docker-related commands on the server where the Docker engine is installed. The application uses the SSH protocol to establish a secure connection with the server and execute the commands. Further, an alternative approach is also explored, which involves connecting the application with the Docker engine using HTTP. This proposed Docker controller application provides a significant advantage for managing Docker resources remotely, which is highly beneficial in DevOps fields. It provides a user-friendly interface to manage containers, making it easy to create, start, stop, restart, and remove containers. It abstracts away the complexities of working with Docker commands, allowing users to interact with containers more intuitively. It can be used to manage a number of docker engines from one place making it easy to control and monitor all the docker resources. Its performance, security, and scalability are evaluated using various testing techniques, and the results are found satisfactory. Further improvements may include enhancing the application's features, optimizing the performance, and exploring other possible approaches for establishing the connection between the application and the Docker engine.

## I. Introduction

Containerization technology is a method of operating system virtualization. It allows the applications with their dependencies to be wrapped together in a portable and isolated environment called the container. Containers provide a consistent and lightweight runtime environment for running applications [1]. In recent years, containerization technology has gained immense acceptance due to its capability to package applications and their dependencies into a single unit. Docker is one such popular containerization technology used widely across the world [2]. However, managing Docker resources on a remote server can be a challenging task. The need for a user-friendly, easy-to-use application to control Docker resources from a mobile device has arisen. The Flutter app development toolkit by Google [3] allows us to build natively compiled mobile, web, and desktop apps from one codebase. Using Flutter would be the best option for these types of scenarios because of its ability to develop cross-platform apps, that is, only a single code base is required for developing iOS, Android, Windows, or even web apps [4].

The objective of this paper is to design and develop a Flutter-based Docker controller application that facilitates simplified container management and centralized control and monitoring of Docker resources. The application will allow users to start, stop, and manage Docker containers, images, and volumes from a mobile device. It manages several Docker engines centrally as well as controls and monitors all the Docker resources.

In this paper, the two approaches are discussed for the development of the application for these types of problems or scenarios. The first approach is to use the HTTP protocol for the connection between the application and the Linux server where the docker engine is installed while the other approach is to use the SSH protocol for the same task. Also, the parallel comparison of both approaches is carried out in this paper. Though the tech stack, as mentioned in this paper, includes

Flutter (for app creation) and Docker (as a container engine) for the sample implementation, these can be replaced with any equivalent, but the roadmap and design implementation will remain the same. This paper also conducts a comparative study between the two approaches namely, via HTTP protocol and SSH protocol.

The paper starts with the introduction of the problem statement and objectives in Section 1. In Section 2, the prerequisite platforms and frameworks used are described. Section 3 surveys the applications and practices of Docker in various sectors. Section 4 presents the design methodology and implementation of the application. Section 5 discusses the experimentation. Last but not least, Section 6 concludes the paper with a discussion of future works.

## II. PLATFORMS AND FRAMEWORKS

### A. Docker

Docker [5] is an open source project and platform for creating, deploying, running, updating, and managing container-based applications. It is a set of products based on platform-as-a-service (PaaS) that make use of OS-level virtualization to provide software in packages termed containers. It offers the ability to bundle and operate an application in a loosely isolated ecosystem known as a container. On a given host, numerous containers can run simultaneously due to the isolation and security. One can run/execute applications without being dependent on what is already set up or installed on the host since containers are small and have everything needed to run them [6]. One may ship code more quickly, standardize application operations, migrate code without interruption, and save money by maximizing resource utilization using Docker [7].

Docker containers can be used as a key building block for creating cutting-edge platforms and applications. With Docker, one can easily create entirely managed platforms for the developers, deploy the code using standardized continuous integration and delivery pipelines, and build a highly scalable system for data processing [7].

### B. Kubernetes

An internal Google project gave rise to Kubernetes [8], an open-source container orchestration technology. Kubernetes plans and automates the tasks necessary for managing container-based systems, including deployment of containers, updates, discovery of services, provisioning of storage, load balancing, and health monitoring, etc. [9]. Additionally, owing to the open-source Kubernetes tool ecosystem, which includes Istio and Knative, any organization can create a high-productive platform-as-a-service (PaaS) for applications based on containers and a shorter on-ramp to the serverless computing. A container orchestration tool is necessary for controlling and monitoring the lifecycles of containers in more complicated systems. Even though Docker has a built-in orchestration tool (named Docker Swarm), Kubernetes is preferred by most developers [9].

### C. SSH and HTTP

A secure channel between two computers over an unsafe network is made possible by the network protocol known as SSH (Secure Shell) [10]. It is widely used to provide secure remote access to servers and network devices. SSH offers authentication, confidentiality, file transfer, and integrity via various cryptographic approaches [11].

HTTP (Hypertext Transfer Protocol) is based on the TCP (Transmission Control Protocol) and it plays a vital role in the modern web ecosystem as an application-layer protocol for communication between web browsers and servers. Web resources can be efficiently retrieved and delivered due to their stateless nature and request-response cycle [12].

### D. Flutter SDK

Flutter [3] is a software development kit for developing mobile applications using Dart, a programming language. It is an open-source user interface (UI) toolkit designed and developed by Google and has emerged as a prominent solution for cross-platform mobile application development. The main ability of Flutter is to develop high-performance apps just by using a single codebase. It has features like hot reload for rapid iterations and a rich set of highly customizable widgets. Flutter is indeed a powerful yet robust solution for the development of iOS and Android platforms based applications. This framework has prominent usability and performance and offers great potential for reducing development time and effort while maintaining an excellent user experience [4].

## III. SURVEYS OF DEVOPS PRACTICES IN MARKET

According to the article on Medium [13], there has been a 50% rise in the use of containers technology by organizations even in 2018. More than 12,000 enterprises use Docker. Most of these businesses are situated in the United States and are involved in software and information technology based services. Additionally, there is a significant chance that Docker clients would possibly use Jenkins and Kubernetes [13][14].

Docker has found widespread adoption and application in various industries. For instance, computer software and information technology services are the largest sectors. Docker is extensively used for software development, testing, and deployment. It enables the developers to package their applications with dependencies into containers, which ensures consistency across various environments. Docker also comes out to be the best option during the creation of micro-services based applications, where Docker can help in breaking down the application into smaller and more manageable components. Docker has gained significant penetration in the IT industry too, with a large number of companies leveraging its benefits. Many organizations have embraced container orchestration platforms like Kubernetes to manage and scale Docker containers effectively. In addition to it, there are many other segments of industries where Docker is used e.g. internet, financial services, marketing and advertising, education, and health care. In healthcare and education, Docker is used to improve the deployment and management of healthcare and teaching applications. It helps ensure consistency and compatibility across different systems, reducing integration challenges [14].

## IV. DESIGN METHODOLOGY AND IMPLEMENTATION

This section provides an overview of the system's requirements, design, and implementation details. The design phase includes listing and designing the required and additional features for the apps, figuring out the proper data flow and networking policies, etc. The system has been implemented using the Flutter framework for building the application interface. The two approaches proposed to design the application are as follows. The choice of approach depends on the particular requirements of the system and the security concerns of the organization. The basic structure and flow of the system is depicted in Fig. 1.

### A. Using HTTP Protocol for the Connection and Data Transfer

The preliminary requirement of the system is that the server where the container engine is installed must be able to provide the HTTP APIs for establishing the proper connection between the application created and the server where the container engine is installed. In this case, the server where the Docker engine is installed will also act as the web server and provide the APIs to run the Docker related commands on the server. The Flutter application will communicate with the server using HTTP requests and responses. This approach requires installing a script file in the Docker server, which may have some security concerns. However, it is faster and more scalable than the SSH approach, making it favorable for large-scale environments. However, the security concerns in this approach bring down scalability when choices are made.

### B. Using SSH Protocol for the Connection and Data Transfer

The preliminary requirement in the case of SSH based connection is that the server where the container engine is installed must be open to accepting SSH requests for establishing the connection between the application and the Docker engine. Moreover, the proper authorization should be given for the execution of the docker related commands in the Linux server. The system establishes a secure SSH connection with the server where the Docker engine is installed to manage Docker resources. The user can perform various Docker-related operations through the interface, and the application sends the corresponding commands over SSH to the Docker engine. This approach ensures security and reliability by establishing a secure connection with the server where the Docker engine is installed. In this case, the main thing is the proper connection establishment via SSH and also the proper authorization method establishment [15].

### C. Detailed Design

The basic sequence diagram for the development of the app is demonstrated in the following Fig. 2.

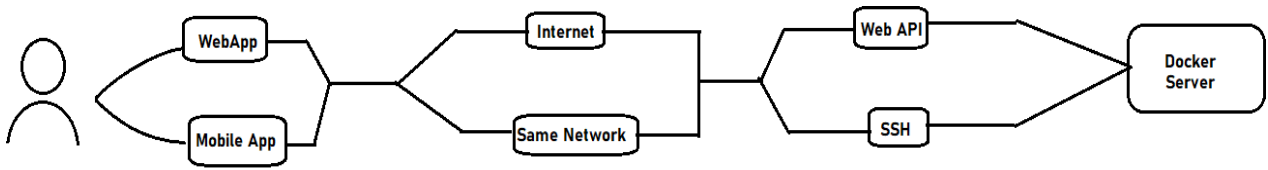

Fig. 1. The basic flow of the system.

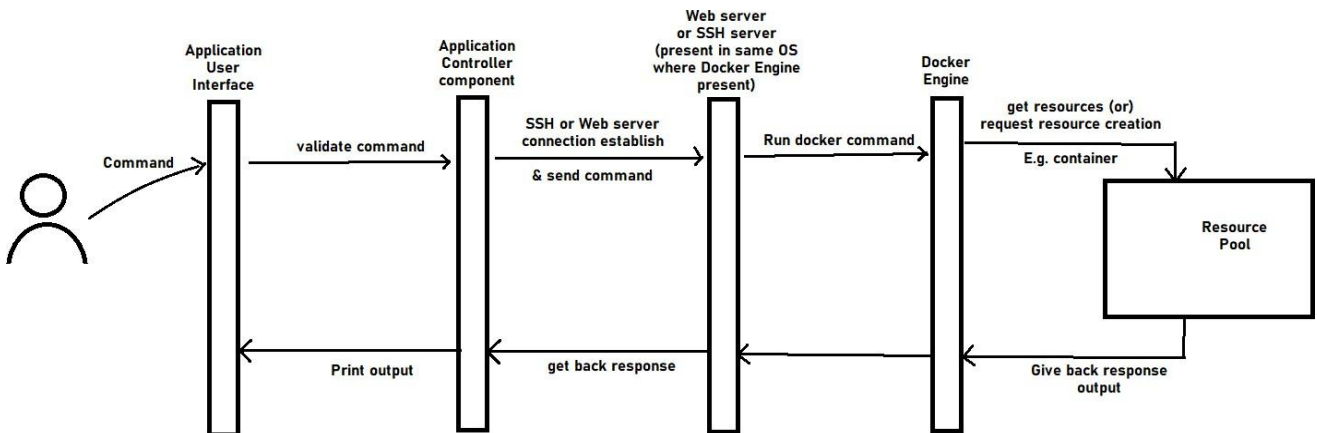

Fig. 2. Sequence diagram.

At first, the user enters the command in the application, and the command is then validated in the application using the validation methods present in the application. The application then connects with the SSH server or HTTP server as the IP address provided to the application at the start of the application. Depending upon the protocol used, a proper authentication should also be done if required. For instance, for the SSH connection, we might need either the user password or the user enters the SSH key for securely logging into the server and running the command. In the same manner, we can implement a similar kind of security system for the HTTP based connection, for instance, requiring a password or key before executing any particular command. On sending the request, the controller component of the application sends the request to the web server or the HTTP server which is present in the same OS where the Docker engine is already set up (in our case). However, the Docker engine can be set up separately as well. The Docker engine is then responsible for running the Docker specific commands and requesting the resources from the resource pool. The output response from the Docker comes back to the application and gets printed in the user interface of the application.

### D. HTTP Oriented Implementation

At first, the initial setup of the project is required which involves setup of the Docker server, web server to accept the requests, and the network configurations like setup of the firewall or opening of the ports for request, etc.

To install the web script on the server that provides API for HTTP connection and runs commands on the Linux server, we can use an HTTPD or even any framework for that. In our case, we have used Apache HTTPD as we just need to demonstrate the working of the application. In the APIs, we can define endpoints that allow clients to send requests containing the commands they want to run on the server. Once we have defined the endpoints, we can use a package like subprocess to execute the commands on the server. The subprocess package allows us to run any command in the Linux server, as in our case, we run the Docker specific commands. After creating the web script, we need to deploy it to the server so that it can serve the request. Once the web script is installed on the server, we can use HTTP requests to execute commands on the server.

We also need to make sure that the server is accessible from the internet or in other cases, it must be present in the same network and has a static IP address or domain name that can be used to connect to it. Next, we need to ensure that the server's firewall is properly configured to allow incoming connections on the relevant ports.

As shown in Fig. 3, on the homepage of the application, we need to enter the IP or the hostname of the server where the docker engine is installed and ready to provide the services. After connecting the device, the device appears in the list of devices. In one instance, we can connect any number of devices and actually manage them from one single application. At present, the application does not include features like user authentication or such parameters.

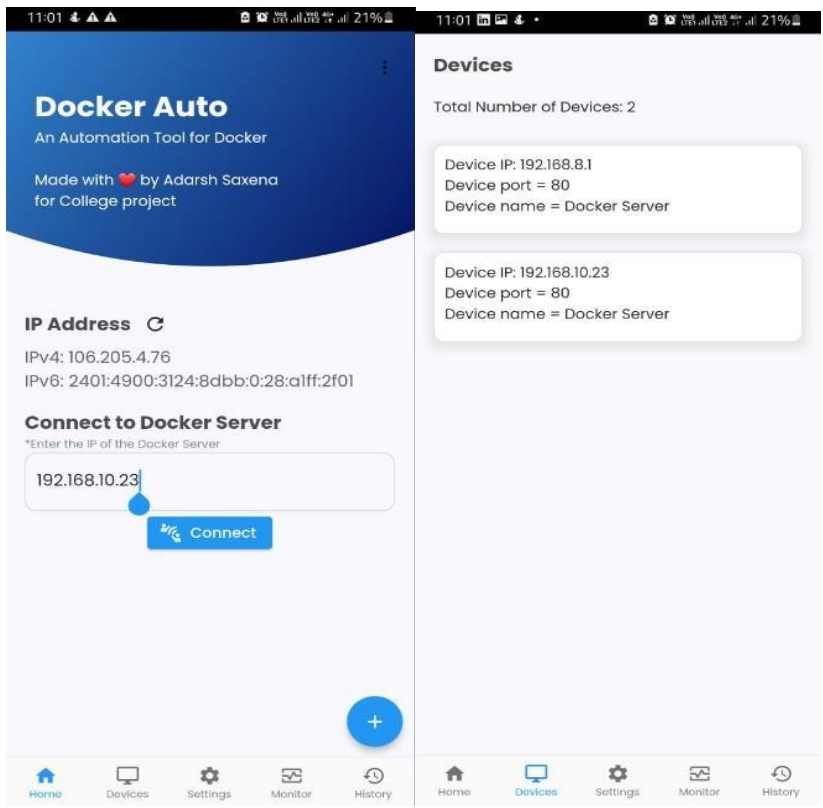
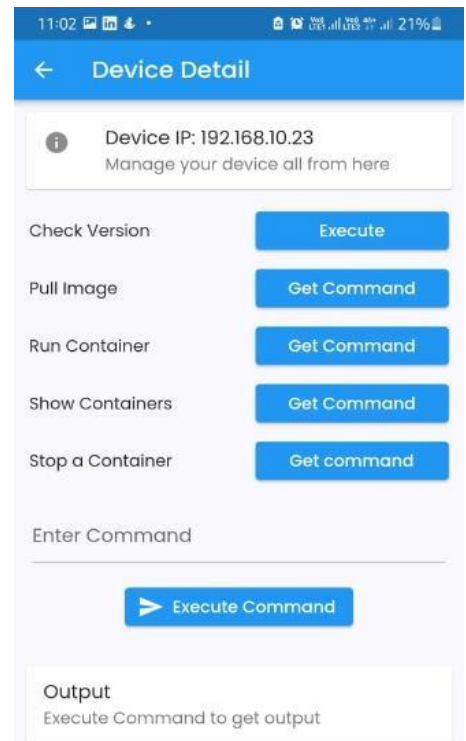

Fig. 3. Homepage of the application and list of devices after connecting them.

Fig. 4. Device specific portal to run commands inside that.

In Fig. 4, the device with IP 192.168.10.23 is open and from here one can manage the docker resources. The commands can be run directly by clicking on the button available for a few tasks while for some other tasks, the specific commands can be customized.

In the Flutter application, for making HTTP requests to the server, we can employ the built-in HTTP package in the Dart programming language. This package offers high-level APIs for making HTTP requests and handling responses, making it easy to interact with our API endpoints. Once we have sent the request, we can handle the response from the server in our Flutter application.

*E. SSH Oriented Implementation*

For establishing the SSH based connection, the SSH2 package is used in the Flutter application. To establish an SSH connection with the server, we first create an instance of the SSH Client class and pass in the host, port, username, and password of the remote server. We can then connect to the server using the connect method and execute commands on the server using the execute method. One thing that needs to be taken care of is that the username of the user which is being used for logging into the server should have all the authorizations for running the Docker related commands into the server. The error handling and exception handling mechanisms are also implemented to ensure that the application remains stable and functional even in the event of unexpected errors or exceptions.

In Fig. 5, three consecutive screens are shown for the application which works on the SSH based protocol for the execution of the Docker related commands. At first, one needs to enter the IP address of the Linux server, where the docker engine is installed and the SSH is also configured. Then, the IP address will be added to the application as a new device instance which can be used to connect to it and execute commands inside that. Next, going inside the Device menu, one can see all the devices that are there in the application database at present. Next, going inside a particular device, one can see the text fields to enter the SSH credentials that are required for logging into the server and then the command to execute.

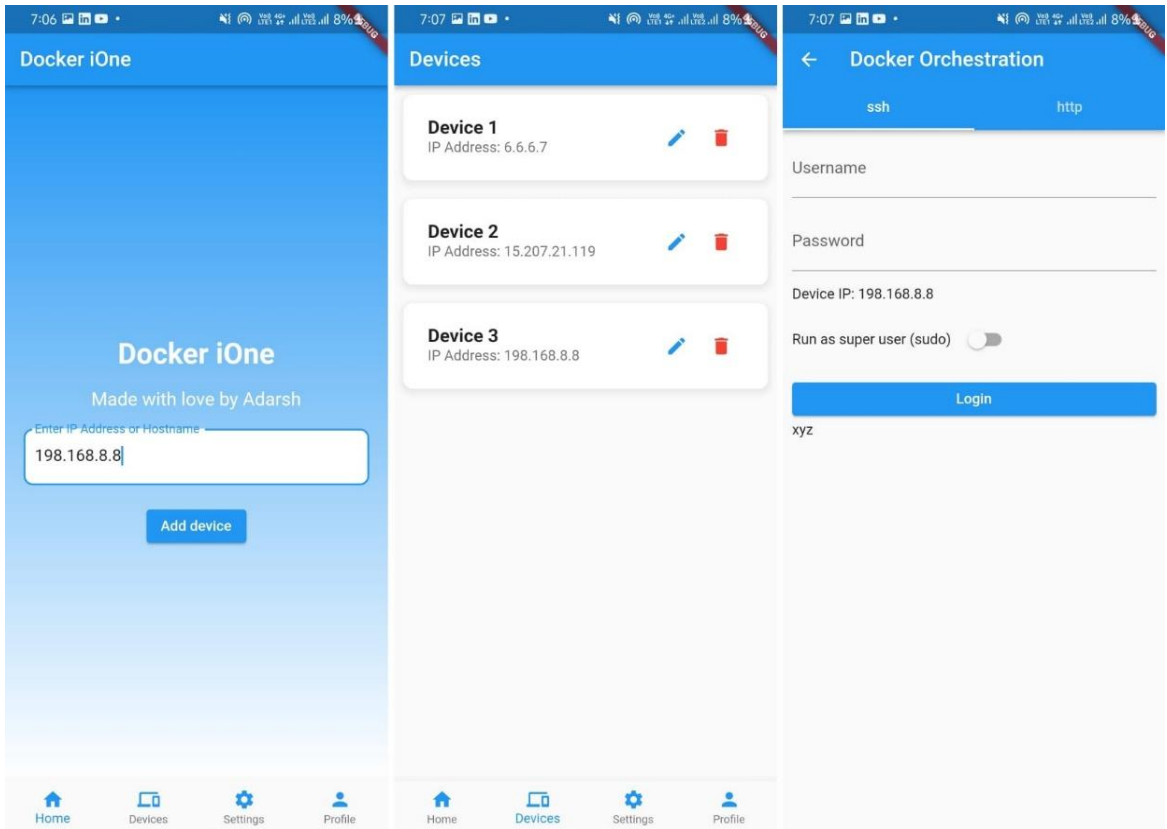

Fig. 5. SSH based Application

## V. EXPERIMENTATION AND DISCUSSION

A server with a Docker engine installed is required to test the application. The server can be a remote server or a local virtual machine running Docker. The major criteria to judge the results of this project include the productiveness that it brings in the implementation of the Docker backed UI, also the user friendliness of the app developed, and the effectiveness of the features present in the application. Moreover, the application portal can be tested based on the security parameters that are generally used in the testing of the applications.

The SSH-based connection can be tested by establishing a connection with the server and sending different commands to the server to check if the server is responding correctly. This is done by simulating different scenarios like invalid credentials, network failure, etc. to ensure that the application is resilient to such scenarios. In most cases, manual testing is done for different components present in the application.

The HTTP-based connection can be tested by sending different API requests to the server and checking the response. This is done on the basis of the same parameters that were present in the SSH-based connection except that in this case, we can use Postman for the API testing.

During the testing process, we also encountered limitations with the use of SSH for communication with the Docker server. While SSH is a secure and widely used method for remote access and communication, it does have some limitations in terms of performance and scalability. As a potential future extension, we could explore the use of other communication protocols such as HTTP with encryption and other security parameter can be implemented. However, in this paper, we are using HTTP protocol without TLS or SSL configured.

Out of the two approaches, namely HTTP-based connection and SSH-based connection for running Docker command, presented in this paper, here is a quick comparison between the two based on different parameters.

*1) Security:* SSH connection provides better security as it uses encrypted communication to establish a connection between the client and server. On the other hand, HTTP-based connection uses unencrypted communication which can pose a security threat if not properly secured. Though, out of the box, TLS or SSL can be configured to provide encryption-in-transit for the connection.

*2) Scalability:* Both SSH and HTTP-based connections can be scaled, but HTTP-based connections are more scalable as they can handle a larger number of requests due to the stateless nature of the HTTP protocol.

*3) Performance:* SSH-based connection provides better performance as it has a lower overhead compared to HTTP-based connections. SSH connections have lower latency and faster response times as compared to HTTP-based connections.

## VI. CONCLUSION AND FUTURE WORKS

In this paper, the Flutter application to control Docker resources has been designed and developed, and all the requirements and specifications have been met. The application enables users to manage their Docker resources efficiently and securely through a user-friendly interface with centralized control and monitoring. The application is also scalable, making it suitable for deployment in large-scale production environments. For HTTP connection oriented applications, the application is more easily scalable as compared to the SSH based connections. Though the SSH based application provides more security compared to the HTTP based counterpart, unless proper encryption and authorization methods are implemented in HTTP based application.

One of the areas of future work includes strengthening the security protocols for data security and privacy. For instance, use of the proper authentication and authorization while logging into the server. A more robust encryption method and performance data can add more credibility. Apart from this, the compatibility of the application can be improved, and the also its integration with different platforms and tools or technologies such as the integration of the application with CI/CD pipeline. Moreover, the integration of the application with the terraform for setting up the Docker engine and server which is in-born compatible with this application would make the adoption of the application much easier. Another area for future work is the improvement of the user interface to enhance the user experience and provide additional functionality, such as the ability to manage container logs or network configurations. The experimentation can also be expanded by considering more empirical data supporting the application's efficacy.

## REFERENCES

[1] O. Bentaleb, A. S. Z. Belloum, A. Sebaa, and A. El-Maouhab, "Containerization technologies: taxonomies, applications and challenges," *J. Supercomput.*, vol. 78, no. 1, pp. 1144–1181, Jan. 2022, doi: 10.1007/S11227-021-03914-1/FIGURES/5.

[2] A. Eftimie and E. Borcoci, "Containerization Using Docker Technology."

[3] "Flutter - Build apps for any screen." https://flutter.dev/ (accessed Oct. 28, 2024).

[4] R. Edge and A. Miola, *Cross-Platform UIs with Flutter: Unlock the ability to create native multiplatform UIs using a single code base with Flutter 3*. Packt Publishing Ltd, 2022.

[5] D. Merkel, "Docker: Lightweight Linux Containers for Consistent Development and Deployment," Accessed: Oct. 28, 2024. [Online]. Available: http://www.docker.io.

[6] "What is Docker? | Docker Docs." https://docs.docker.com/get-started/docker-overview/ (accessed Oct. 28, 2024).

[7] "What is Docker? | AWS." https://aws.amazon.com/docker/ (accessed Oct. 28, 2024).

[8] "Overview | Kubernetes." https://kubernetes.io/docs/concepts/overview/ (accessed Oct. 28, 2024).

[9] "What Is Docker? | IBM." https://www.ibm.com/topics/docker (accessed Oct. 28, 2024).

[10] "What is SSH (Secure Shell)? | SSH Academy." https://www.ssh.com/academy/ssh (accessed Oct. 28, 2024).

[11] "OpenSSH." https://www.openssh.com/ (accessed Oct. 28, 2024).

[12] "Architectural Styles and the Design of Network-based Software Architectures." https://ics.uci.edu/~fielding/pubs/dissertation/top.htm (accessed Oct. 28, 2024).

[13] "Interesting facts — Companies and the use of Docker | by Level Up Education | Medium." https://medium.com/@tao_66792/interesting-facts-companies-and-the-use-of-docker-948baa8cf309 (accessed Oct. 28, 2024).

[14] "Companies using Docker and its marketshare." https://enlyft.com/tech/products/docker (accessed Oct. 28, 2024).

[15] A. H. Al-Hamami and G. M. W. Al-Saadoon, "Handbook of research on threat detection and countermeasures in network security," *Handb. Res. Threat Detect. Countermeas. Netw. Secur.*, pp. 1–426, Oct. 2014, doi: 10.4018/978-1-4666-6583-5.